\documentclass[11 pt]{article}%
\usepackage{amsmath}%
\usepackage{amsfonts}%
\usepackage{amssymb}%
\usepackage{epsf}%
\usepackage{hyperref}%

\def\singlespace {\smallskipamount=3.75pt plus1pt minus1pt
                  \medskipamount=7.5pt plus2pt minus2pt
                  \bigskipamount=15pt plus4pt minus4pt
                  \normalbaselineskip=15pt plus0pt minus0pt
                  \normallineskip=1pt
                  \normallineskiplimit=0pt
                  \jot=3.75pt
                  {\def\smallskip {\vskip\smallskipamount}}
                  {\def\medskip   {\vskip\medskipamount}}
                  {\def\bigskip   {\vskip\bigskipamount}}
                  {\setbox\strutbox=\hbox{\vrule
                    height10.5pt depth4.5pt width 0pt}}
                  \parskip 7.5pt
                  \normalbaselines}
\def\middlespace {\smallskipamount=5.625pt plus1.5pt minus1.5pt
                  \medskipamount=11.25pt plus3pt minus3pt
                  \bigskipamount=22.5pt plus6pt minus6pt
                  \normalbaselineskip=22.5pt plus0pt minus0pt
                  \normallineskip=1pt
                  \normallineskiplimit=0pt
                  \jot=5.625pt
                  {\def\smallskip {\vskip\smallskipamount}}
                  {\def\medskip   {\vskip\medskipamount}}
                  {\def\bigskip   {\vskip\bigskipamount}}
                  {\setbox\strutbox=\hbox{\vrule
                    height15.75pt depth6.75pt width 0pt}}
                  \parskip 11.25pt
                  \normalbaselines}
\def\doublespace {\smallskipamount=7.5pt plus2pt minus2pt
                  \medskipamount=15pt plus4pt minus4pt
                  \bigskipamount=30pt plus8pt minus8pt
                  \normalbaselineskip=30pt plus0pt minus0pt
                  \normallineskip=2pt
                  \normallineskiplimit=0pt
                  \jot=7.5pt
                  {\def\smallskip {\vskip\smallskipamount}}
                  {\def\medskip   {\vskip\medskipamount}}
                  {\def\bigskip   {\vskip\bigskipamount}}
                  {\setbox\strutbox=\hbox{\vrule
                    height21.0pt depth9.0pt width 0pt}}
                  \parskip 15.0pt
                  \normalbaselines}

\begin{document}
\begin{center}
{\bf {\Large Particle creation in (2+1) circular dust collapse}}
\bigskip

{\large Sashideep Gutti \footnote{e-mail address: sashideep@tifr.res.in}
and 
T. P. Singh \footnote{e-mail address: tpsingh@tifr.res.in},

}
\bigskip

{\it Tata Institute of Fundamental Research,}\\
{\it Homi Bhabha Road, Mumbai 400 005, India}
\medskip

\end{center}
\bigskip
\bigskip

\begin{abstract}

\noindent We investigate the quantum particle creation during the circularly
symmetric collapse of a 2+1 dust cloud, for the cases when the cosmological
constant is either zero or negative. We derive the Ford-Parker formula for
the 2+1 case, which can be used to compute the radiated quantum flux in the
geometric optics approximation. It is shown that no particles are created
when the collapse ends in a naked singularity, unlike in the 3+1 case. When
the collapse ends in a BTZ black hole, we recover the expected Hawking
radiation.  

\end{abstract}

\section{Introduction}
There are many analytical examples of formation of black holes and naked 
singularities, in the classical theory of general relativity \cite{si1}. In 
order to
study physical phenomena in the vicinity of the spacetime singularity, where 
the curvature is extremely high, one must take into account quantum effects 
such as particle creation, as also the effects resulting from  quantization of 
the gravitational field. In the case of a black hole, the presence of the
event horizon makes the study relatively easier, and particle creation is
described by the well-known Hawking radiation, except possibly in the 
end-stages of black hole evaporation, where quantum gravitational effects can
modify physics in a manner that is not yet properly understood.  

In the investigation of quantum particle creation in the vicinity of a naked
singularity, one cannot apply standard methods from black hole physics, such
as Bogoliubov transformations, because the presence of a Cauchy horizon 
leads to a breakdown of predictability, and the future null infinity in the
spacetime is not well-defined \cite{si2}. Nonetheless useful information about
the energy associated with the quantum particle creation in the approach
to a naked singularity can be obtained through approximate methods. One such
approximation, developed by Ford and Parker \cite{ford}, is to calculate the 
radiated power, in the geometric optics approximation,  when one quantizes a 
massless scalar field on the background of a collapsing spherical body.
Another useful method is to calculate the vacuum expectation value of the
two dimensional stress-energy tensor of a quantized scalar field, by suppressing 
the angular coordinates \cite{hiscock}, in a collapsing spherical geometry.
In more recent times, the study of quantum mechanics of naked singularities was
heralded by Vaz and Witten \cite{vaz1} \cite{vaz2}.

In a series of papers \cite{s1}, \cite{s2}, \cite{s3}, \cite{s4}, \cite{s5} 
we have applied both these methods to
study quantum particle creation in the spherical four dimensional collapse
of a pressureless inhomogeneous dust cloud, described by the 
Lemaitre-Tolman-Bondi (LTB) model. The nature of singularities in classical 
LTB collapse, and their dependence on initial data, has been studied in great 
detail \cite{sj1}, \cite{sj2}, \cite{sj3}. For some initial data, the LTB 
collapse ends in a black hole, and for other (generic) initial data it ends in
a naked singularity. When the LTB collapse ends in a black hole, the quantum
particle creation is described by Hawking radiation, as expected.

However, when the collapse ends in a naked singularity, the nature of
particle creation is very different - it is typically found that in the
approach to the Cauchy horizon the integrated quantum flux diverges. This
signals a breakdown of the semiclassical approximation and the need
for a full quantum gravitational treatment, as the Cauchy horizon is
approached. This conclusion is reaffirmed by the result that if the 
semiclasical analysis is terminated at the epoch when curvatures reach Planck 
scale, the integrated flux is only of the order of one Planck unit
\cite{six}.

Subsequently, 
in order to study the role of quantum gravity effects in
LTB collapse, we set up a midisuperspace quantization scheme, in the
framework of canonical quantum general relativity \cite{wd},
following earlier pioneering work by Vaz and Witten \cite{vawi1}, \cite{vawi2}.
A Wheeler-DeWitt
equation for describing the quantum gravitational collapse of a spherical dust 
cloud, matched to a Schwarzschild exterior, was set up. In order to 
arrive at a solution of the Wheeler-Dewitt equation, it is necessary to 
implement a regularization, and we investigated the consequences of a 
lattice regularization. We could show how Hawking radiation arises as an
approximation to quantum gravity, when the classical collapse results in a 
black hole \cite{k1}, \cite{k2}, \cite{k3}, \cite{k4}. However, as things
stand, it is not clear that the lattice regularization is well-suited to
address the behavior of the collapse near the curvature singularity, in
particular near the naked singularity. 

Considering the difficulty encountered in finding a sufficiently general 
and useful regularization scheme for the midisuperspace Wheeler-DeWitt 
equation in the 3+1 case, it was felt desirable to adapt the above 
quantization programme to the lower dimensional 2+1 case of circularly 
symmetric dust collapse. In the first of this series of papers on
2+1 dust collapse \cite{sashi}, classical inhomogeneous dust collapse
was studied. For the relatively less interesting case of a flat exterior
spacetime with a zero cosmological constant, it was found that the collapse
ends in a naked singularity. When a negative cosmological constant is present
and the exterior is an AdS spacetime, then for certain initial conditions 
the collapse ends in the well-known BTZ black-hole \cite{banados}, whereas for
other initial conditions it ends in a naked singularity.

In the present paper, which is the second in this series on quantization of
lower dimensional collapse, we will apply the methods of Ford and Parker to
calculate the quantum particle creation, both in the black hole case and in the naked
singularity case, with and without the cosmological constant. This study 
serves as a prelude to the full quantum gravitational treatment of the 2+1
collapse using canonical methods, which will be taken up subsequently. The
unusual structure of the spacetime infinity in the AdS case necessitates a 
careful treatment of the quantum field theoretic problem on an AdS 
background, (in particular the definition of an initial quantum vacuum), 
which has been undertaken in \cite{isham}. Following the work of 
\cite{balasubramaniyan} and
\cite{vakkuri} we will assume reflecting boundary conditions at spacelike 
infinity. Hawking radiation in 2+1 collapse has been studied also by 
\cite{korea}.

The plan of this paper is as follows. In Section 2 we recall the key results 
from \cite{sashi}, on classical 2+1 dust collapse. In order to study particle
creation in the geometric optics approximation, the Ford-Parker formula for
the case of a zero cosmological constant is worked out in Section 3. In Section
4 this formula is used to show that no particle creation takes place
when the collapse ends in a naked singularity. One of the key results of this
paper is given in Section 5, where we derive, from first principles, the 
Ford-Parker formula on an AdS background. In Section 6, 7 and 8,
 this formula is used
to show that while the BTZ black hole resulting from collapse emits Hawking
radiation, the collapse ending in a naked singularity on an AdS background
again gives rise to no quantum particle creation.  

It is important to ask why one should investigate gravitational collapse in 
the apparently `physically unrealistic' 2+1 dimensional case, considering 
that the observed Universe has at least four spacetime dimensions. Is such a
study carried out only because it is simpler than the four dimensional case,
with the simplification having been bought at the expense of introducing
`physical unrealism'?  The answer is no. One can definitely learn
useful physics about gravitational collapse in four or higher dimensions, by 
examining the corresponding situation in 2+1 dimensions. This is because 
there are no gravitational waves in 2+1 gravity, since the Weyl tensor here
is exactly zero. Yet, gravity in a 2+1 spacetime is non-trivial; it admits the
formation of the BTZ black hole, and also of a naked singularity. The 
absence of gravitational waves holds out the promise that one can build an
exact model of a quantum black hole in 2+1 dimensions, something that may
well be impossible in complete generality in
a 3+1 quantum gravity. Such a 2+1 model may well possess
features (such as counting of microstates that lead to the Bekenstein-Hawking
entropy) which are universal and independent of spacetime dimensions; but
this is something which we can ascertain only after we have built such a model.
It is just that the presence of a non-zero Weyl tensor in four or higher 
dimensions complicates attempts to construct a general exact model of a quantum
black hole. Further, the 2+1 BTZ black hole is the simplest non-trivial
and useful black hole system. (An exact solution with a black hole might exist
in 1+1 dimensions \cite{CGHS}, but since there is only one space dimension, 
the horizon of the black hole consists of only two points and hence one 
cannot meaningfully talk of the area of a black hole horizon in this case).

Furthermore, one could question the introduction of a negative cosmological
constant, as is done in this paper, when the observed Universe has  a
cosmological constant which is perhaps positive, or at best zero, but certainly
not negative. The reason is that in 2+1 dimensions a black hole solution 
arises only when the cosmological constant is negative. When it is zero or 
positive, the gravitational collapse necessarily ends in a naked singularity. 
Thus if we want to build an exact quantum model of the BTZ black hole, we 
must restrict to the case of a negative cosmological constant. There are
also reasons to believe that it would not make sense to directly construct a 
quantum black hole model in a higher dimensional space with a positive 
cosmological constant, because quantum gravity in such a spacetime may not 
exist nonperturbatively \cite{GKS}, \cite{Witten}. Pure 
quantum gravity with a positive cosmological constant may hence not exist 
as an exact theory, but only as a part of a larger system \cite{Witten}.

Thus, for the reasons described in the previous two paragraphs, it is useful
to investigate semiclassical gravitational collapse in 2+1 dimensions, as is 
done in this paper, as a prelude to constructing a model of a quantum 
gravitational black hole.

\section{Inhomogeneous dust collapse in (2+1) dimensions}
We summarize here the classical collapse solution for inhomogeneous dust collapse for the cases with zero and negative cosmological constant, analyzed   in \cite{sashi}. The collapse of homogeneous dust in (2+1) dimensions was first studied by \cite{mann1}. This was followed by other interesting work in (2+1) dimensions, including shell collapse, \cite{mann2}, \cite{peleg} and other scenarios\cite{barrow},\cite{hortacsu1},\cite{hortacsu2}. We consider circularly symmetric dust and set up a comoving and synchronous coordinate system
\begin{equation}
ds^{2}=-dt^{2} + e^{2b(t,r)}dr^{2}+R(t,r)^2d\phi^2
\label{3dmetref}
\end{equation}
where $t$ is the comoving time, $r$ is the shell label and $\phi$ is the angular coordinate. We evaluate the Einstein equations and solve for the metric components. For the case with zero cosmological constant the metric (\ref{3dmetref}) becomes
\begin{equation}
ds^2=-dt^2 +\frac{(c_1't+1)^2dr^2}{c_1^2-2\kappa\int_{0}^{r}{\rho _i(s) sds}+1}+(c_1t+r)^2d\phi^2
\label{0metricref} 
\end{equation}
 where $c_1$ is a function of $r$ alone and ' denotes derivative w.r.t $r$. $R=c_1(r)t+r$ has the interpretation of being the physical radius of a shell with label $r$. $c_1$ is interpreted as the initial velocity of the shell with label $r$. The function $\rho_i(s)$ is the initial density profile of the dust. A curvature singularity develops when the physical radius of a shell becomes zero. A shell becomes singular only if the initial velocity  $c_1$ is negative. The time for singularity formation for a given shell is given by  $t=-r/c_1$. The analysis of outgoing null rays show that the singularity is always locally naked.
In (2+1) dimensions the circularly symmetric vacuum solution for the case of zero cosmological constant is a flat conical spacetime \cite{peleg} of the form
\begin{equation}
ds^2=-dT^2+dR^2+\alpha^2 R^2d\phi^2
\label{fordqexterior}
\end{equation}
where $T$ is the time coordinate, $R$ is the radial coordinate and $\phi$ is the angle, with the deficit angle given by $2\pi (1-\alpha)$ when $\phi$ goes from $0$ to $2\pi$.

 The spacetime exterior to the collapsing dust cloud is given by Eqn. (\ref{fordqexterior}). Matching the first and second fundamental forms across the outer boundary ($r_b$) of the collapsing cloud yields the value  
\begin{equation}
\alpha=\sqrt{1-2\kappa\int_{0}^{r_b}{\rho _i(s) sds}}
\label{alpharef}
\end{equation}

Solving the Einstein equations with a  negative cosmological constant case yields the following metric,
\begin{equation}
ds^2=-dt^2+\frac{(\cos(\sqrt{\Lambda }t)+B'\sin(\sqrt{\Lambda }t))^2dr^2}{\Lambda r^2+\Lambda B^2-2\kappa\int_{0}^{r}{\rho _i(s) sds}+1}+(r\cos(\sqrt{\Lambda }t)+B\sin(\sqrt{\Lambda }t))^2d\phi^2
\label{nmetricref}
\end{equation}
where $\Lambda$ is the absolute value of the cosmological constant.
The function $B(r)$ defines the initial velocity profile of the dust cloud. The physical radius is 
given by $R=r\cos(\sqrt{\Lambda }t)+B\sin(\sqrt{\Lambda }t)$.
 A shell with label $r$ becomes singular when the physical radius $R$ shrinks to zero size. The shell becomes singular at the time $t_s=\arctan{(-r/B)}/\sqrt{\Lambda}$. 
The nature of the singularity can be deduced by analyzing the outgoing null rays. It is shown in \cite{sashi} that the outgoing null rays can emerge from the singularity till the time when the critical shell becomes singular. The critical shell is defined as the shell $r_c$ for which $\int_{0}^{r_c}{\rho _i(s) sds}=1/\kappa$. This implies that during the collapse, the singularity will be at least locally naked till the time when the critical shell becomes singular.
The singularity is therefore timelike for the shells $r<r_c$. It becomes null for $r=r_c$ and is spacelike for $r>r_c$. 
The exterior spacetime in the case of a negative cosmological constant in
(2+1) dimensions is given by the BTZ metric \cite{banados} with zero angular momentum,
\begin{equation}
ds^2=-(\Lambda R^2-M)dT^2+\frac{dR^2}{(\Lambda R^2-M)}+R^2d\phi^2.
\label{nexteriorref}
\end{equation}
Matching the first and second fundamental forms across the boundary of the cloud yields 
\begin{equation}
M=2\kappa\int_{0}^{r_b}{\rho _i(s) sds}-1.
\label{nMref}
\end{equation}
As demonstrated in \cite{sashi},\cite{mann1}, when $M$ is negative, the collapse ends in
a naked singularity, whereas it ends in a BTZ black-hole when $M$ is 
positive. For suitable initial conditions the naked singularity is globally
naked - we will show below that in such a case no quantum particle creation
takes place.  

\section {Ford-Parker formula for zero cosmological constant}
If one considers the quantization of a massless scalar field on the 
classical gravitational background of a collapsing spherical object, the radiated flux can be calculated
in the geometric optics approximation, using point-splitting regularization \cite{ford}. This calculation
yields the correct Hawking flux for a black hole, and is especially useful for computing the quantum
flux in the approach to the Cauchy horizon, when the background collapse terminates in a naked 
singularity. The key input in the calculation is the map from ingoing null rays coming from 
$\mathcal{I}^-$ to outgoing null rays arriving at $\mathcal{I}^+$. In the present Section we will adapt
the Ford-Parker formula to the case of the 2+1 circular collapse with zero cosmological constant.
The 2+1 Ford-Parker formula in an AdS exterior background is derived in Section 5. 

The exterior spacetime metric is given by (\ref{fordqexterior}). A massless scalar field $\Phi$, propagating in the background geometry satisfies the equation $\Box \Phi=0$. We analyze the mode solutions when $R$ goes to infinity. We assume that the scalar field is purely incoming on $\mathcal{I}^-$. The radial ingoing modes pass through the collapsing cloud, get reflected about the point $r=0$ and reemerge as outgoing modes. The ingoing null ray $v=t+r$ on $\mathcal{I}^-$ gets mapped to $u=F(v)$ where $u=t-r$ on $\mathcal{I}^+$. Retracing back from $\mathcal{I}^+$  to $\mathcal{I}^-$ gives a mapping $v=G(u)$. Given a map between the ingoing and outgoing modes ($G(u)$), the particle flux can be calculated in the asymptotic region using the method of Ford and Parker \cite{ford}.

Let $f_{\omega m}$ be the solution of the massless scalar wave equation with the form given by
\begin{equation}
f_{\omega m}=\frac{e^{-im\phi}}{\sqrt{2\pi \alpha}}\left( \frac{e^{-i\omega v}-e^{-i\omega G(u)}}{\sqrt{r}\sqrt{4\pi \omega}}\right).  
\label{fordasy}
\end{equation}
The scalar field $\Phi$ can be expanded in terms of negative and positive frequency modes given by
\begin{equation}
\Phi=\sum_m\int\limits_0^{\infty}d\omega(a_{\omega m}f_{\omega m}+a_{\omega m}^{\dagger}\bar{f}_{\omega m})
\label{fordexpand}
\end{equation}
which corresponds to an outgoing plane wave on $\mathcal{I}^+$. It is 
normalized using the scalar product defined by
\begin{equation}
(f,h)=-i\int_{\Sigma}d\Sigma^{\mu}\sqrt{g_\Sigma}[f(\partial_\mu h^{*})-(\partial_\mu f)h^{*}] 
\label{scalarproduct}
\end{equation}
where $\Sigma$ is a spacelike hypersurface. This implies the normalization,
\begin{equation}
(f_{\omega m},f_{\omega' m'})=\delta(\omega-\omega ')\delta_{mm'}.
\label{normal}
\end{equation}
The $a_{\omega m}, a_{\omega m}^{\dagger}$ have the interpretation of creation and annihilation operators respectively.
The energy-momentum tensor for the scalar field is given by the expression
\begin{equation}
T_{\mu \nu}=\Phi_{,\mu}\Phi_{,\nu}-\frac{1}{2}g_{\mu \nu}\Phi_{,\alpha} \Phi^{,\alpha}.
\label{energymomentum}
\end{equation}
The energy radiated to $\mathcal{I}^+$ can be calculated from the expectation value of the energy momentum tensor, $<0|T_{RT}|0>$ where the component $T_{RT}$ is given by the expression
\begin{equation}
T_{RT}=\frac{1}{2}(\Phi_{,T}\Phi_{,R}+\Phi_{,R}\Phi_{,T}).
\label{Trt}
\end{equation}
This operator is ill-defined because it is quadratic in fields given at the same spacetime point. The expectation value $<0|T_{RT}|0>$ is evaluated using the point-splitting regularization, as in \cite{ford}. The leading order for the energy radiated to $\mathcal{I}^+$ works out to be the same as the expression obtained by \cite{ford} in the four dimensional case, except for the factor of $1/{4\pi R}$. We obtain,
\begin{equation}
<0|T_{RT}|0>=\frac{1}{(4\pi R)(2\pi \alpha)}\sum_m\left|e^{-im\phi}\right|\left[\frac{1}{4}\left(\frac{G''}{G'}\right)^2-\frac{1}{6}\frac{G'''}{G'}\right]
\label{fpformula}
\end{equation}
where a prime on $G$ denotes derivative with respect to $u$.
The power radiated across a disk of radius $R$ in the asymptotic region is given by integrating over the angle $\phi$
\begin{equation}
P=\int<0|T_{RT}|0>Rd\phi=\sum_mP_m(u)=\frac{1}{4\pi }\sum_m\left[\frac{1}{4}\left(\frac{G''}{G'}\right)^2-\frac{1}{6}\frac{G'''}{G'}\right]
\label{fpformulaP}
\end{equation}
where $P_m(u)$ is the power radiated for the mode $m$.

\section{Particle creation : case with zero cosmological constant}
In order to apply the Ford-Parker formula (\ref{fpformulaP}) for the case
with zero cosmological constant, we now calculate the map $G(v)$ using the
classical solution summarized in Section 2.    
As noted above, for zero cosmological constant the collapse ends in a naked
singularity.
A radial null ray from $\mathcal{I}^-$ in the exterior passes through the dust cloud, goes through the center and reemerges as an outgoing null ray which escapes to $\mathcal{I}^+$.  To evaluate the quantum particle creation by the singularity, we note that the non-singular spacetime terminates on the Cauchy horizon. We then obtain the map between the ingoing and the outgoing null rays in the neighborhood of the Cauchy horizon, and calculate the quantum flux using the equation (\ref{fpformulaP}).

The conical exterior spacetime is given by (\ref{fordqexterior}). The  radial null rays in the exterior are given by $V=T+R$ and $U=T-R$. These are obtained by using the null condition on the metric. In the interior of the dust cloud, the metric is given by (\ref{0metricref}). The expressions for the radial null rays are obtained by imposing the null condition on the metric. This gives the following equation
\begin{equation}
\frac{dt}{dr}=\pm\frac{(c_1't+1)}{\sqrt{c_1^2-2\kappa\int_{0}^{r}{\rho _i(s) sds}+1}}.
\label{q0nullg}
\end{equation}

We can integrate the equation and get $v=t+F_-(r)$ and $u=t-F_+(r)$ where
\begin{equation}
F_\pm= \left[e^{\pm\int{\frac{c_1'dr}{\sqrt{c_1^2-2\kappa\int_{0}^{r}{\rho _i(s) sds}+1}}}}\right]\left[\int{\frac{e^{\mp\int{\frac{c_1'dr}{\sqrt{c_1^2-2\kappa\int_{0}^{r}{\rho _i(s) sds}+1}}}}}{\sqrt{c_1^2-2\kappa\int_{0}^{r}{\rho _i(s) sds}+1}}+b}\right].
\label{egeod}
\end{equation}
Now the functions $F_\pm(r)$ are differentiable in $r$. (They are independent of time and if they are not well behaved for some $r$ it implies there is a problem defining the null geodesics at all times at that $r$. We exclude such possibilities). The function $c_1^2-2\kappa\int_{0}^{r}{\rho _i(s) sds}+1$ should be positive for the metric (\ref{0metricref}) to be well defined.

The singularity is first formed at $t=0,\ r=0$. The first null ray which escapes from the singularity is the Cauchy horizon. The ingoing null ray which passes through the coordinate $(0,0)$ is given by 
\begin{equation}
v_0=t+F_-(r)=0+F_-(0)=F_-(0).
\label{v_0}
\end{equation}
 This null ray becomes an outgoing null ray which passes through the same point $(0,0)$. The outgoing null ray is given by
\begin{equation}
u_0=t-F_+(r)=0-F_+(0)=-F_+(0).
\label{u_0}
\end{equation}
This null ray forms the Cauchy horizon. To find out the map between the null rays $V_0$ (defined in the exterior, which becomes $v_0$ in the interior) and $U_0$ (the ray in the exterior which is formed when the interior null ray $u_0$ reaches the outer boundary of the collapsing cloud).
Let the outer boundary  of the dust cloud be denoted by $r_b$. The interior time $t_i$ when the null ray $v_0=t+F_-(r)=F_-(0)$ was at $r_b$ is given by $t_i=F_-(0)-F_-(r_b)$.
Now the exterior time is given in terms of the time in the interior cloud as
\begin{equation}
T=t\sqrt{1+c_1^2(r_b)}.
\label{qdteqn2}
\end{equation}
At the boundary $r_b$, the null ray $V$ becomes $v$. The map is linear of the form $V=av+b$, where $a,b$ are constants. The reasoning is as  follows.
The physical radius $R(t,r_b)$ is $c_1(r_b)t+r_b$. So the ingoing exterior null ray $V=T+R$ is
\begin{equation}
V=t\sqrt{1+c_1^2(r_b)}+c_1(r_b)t+r_b= kt+r_b
\label{qdteqn3}
\end{equation}
where $k$ is a constant. Similarly at the boundary the interior null ray is $v=t+F_-(r_b)$. Hence the map is of the form $V=av+b$. 
The outgoing null ray $u$ in the interior is given by $u=t-F_+(r_b)$ at the outer boundary. By a similar analysis, it can be shown that the map between the interior outgoing rays and the exterior outgoing rays is linear, of the form $U=cu+d$. The map between the ingoing and outgoing modes at $r=0$ is $v=u$. So since all the relations are linear, the map between exterior $V$ and $U$ is linear. Hence the map $V=G(U)$ is linear.

To analyze the particle creation due to the singularity, we use the formula (\ref{fpformulaP}). When the function is linear, the particle flux is zero since the formula involves second and third derivatives of $G(U)$ with respect to $U$. This is to be contrasted with the (3+1) case \cite{s2}, \cite{s4} where a similar analysis in the neighborhood of the singularity yields an infinite flux of particles. The present result is not unexpected, considering that the 2+1 exterior 
(for zero cosmological constant) is conical spacetime.

\section{The Ford-Parker formula for AdS background}
The AdS spacetime can be obtained by embedding a hyperboloid $ -U^2-V^2+X^2+Y^2=-\Lambda^2$ in a spacetime which has a metric given by $ds^2=-dU^2-dV^2+dX^2+dY^2$. The induced metric expressed in global coordinates (covering AdS) is 
\cite{hawking}
\begin{equation}
ds^2=\sec^2\rho(-dt^2 +d\rho^2)+ \tan^2\rho d\phi^2.
\label{Aglobal}
\end{equation}
The BTZ spacetime can be constructed by dividing the hyperboloid into three regions \cite{balasubramaniyan} and making one of the coordinates periodic. The BTZ spacetime can be considered a patch of the AdS spacetime covered by the global coordinates. A quantum vacuum can be constructed in global coordinates and can be expressed in terms of modes of the BTZ blackhole to derive the Hawking radiation. In this section we derive the radiated flux of a massless scalar field in an AdS spacetime. 
The particle creation in AdS  blackholes of arbitrary dimension was studied in \cite{vakkuri}. Following 
their approach, we compute a way of deriving radiation from the map of incoming and outgoing null rays in the asymptotic region. We assume that in the far past, the AdS spacetime is in a vacuum state constructed using the global AdS coordinates. The normalizable modes for the global AdS coordinates can be found in \cite{balasubramaniyan}. We consider a massless scalar field propagating in the AdS background. We adopt the reflecting boundary condition at spatial infinity. In the AdS case this is necessary since the spatial infinity is timelike. There is no Cauchy surface for the AdS case since the timelike nature of the 
spatial infinity allows information to enter through the spatial infinity. So we assume a reflecting boundary condition. The Klein Gordon equation in the global coordinates is give by $\Box\Phi=0$, written explicitly as
\begin{equation}
\frac{1}{\sec^2\rho}\left(-\frac{\partial^2\Phi}{\partial t^2}+\frac{\partial^2\Phi}{\partial \rho^2}\right)+\frac{1}{\tan\rho}\frac{\partial\Phi}{\partial \rho}+\frac{1}{\tan^2\rho}\frac{\partial^2\Phi}{\partial \phi^2}=0.
\label{Alaplacian}
\end{equation}
We assume the solution to be of the variables separable type $(\Phi=\Phi_t\Phi_\rho \Phi_\phi)$. We 
find the spherical wave solution by putting
${\partial^2\Phi_{\phi}}/{\partial \phi^2}=0$. The equation for the time dependent part is 
$ {\partial^2\Phi_t}/{\partial t^2}+\omega^2\Phi_t=0$. The equation for $\Phi_\rho$ becomes
\begin{equation}
\frac{\partial^2\Phi_\rho}{\partial \rho^2}+\frac{\sec^2\rho}{\tan\rho}\frac{\partial\Phi_\rho}{\partial \rho}+\omega^2\Phi_\rho=0.
\label{Aphirho}
\end{equation}
We change the variables to $x=\cos(2\rho)$, to bring this equation to a known form called the Jacobi differential equation.  Then the equation becomes
\begin{equation}
(1-x^2)\Phi^{''}_{\rho}-(1+x)\Phi^{'}_{\rho}+\frac{\omega^2}{4}\Phi_\rho=0.
\label{Acos2rho}
\end{equation}
Now comparing the above equation to the standard Jacobi differential equation \cite{table} which is
\begin{equation}
(1-x^2)\Phi^{''}_{\rho}-(\beta-\alpha-(\alpha+\beta+2)x)\Phi^{'}_{\rho}+n(n+\alpha+\beta+1)=0
\label{Ajacobdiffn}
\end{equation}
we get $\alpha=0$, $\beta=-1$ and $n=\omega/2$. This implies that $\omega$ cannot have arbitrary values.
The solution is convergent if $\omega$ is an even integer. The solution is the Jacobi polynomial $P^{(\alpha\beta)}_n(\cos2\rho)$. So the positive frequency solution is of the form
\begin{equation}
\Phi_n=e^{-i\omega_n t}P^{(0,-1)}_n(\cos2\rho).
\label{Aphin}
\end{equation}
The high frequency limit for Jacobi polynomials is given by \cite{table}
\begin{equation}
P^{(\alpha\beta)}_n(\cos(\theta)=\frac{\cos{(n+\frac{1}{2}(\alpha+\beta+1))\theta-\frac{\alpha}{2}-\frac{\pi}{4})}}{\sqrt{\pi n}(\sin(\theta/2))^{\alpha+1/2}\cos(\theta/2))^{\beta+1/2}}.
\label{Ahifilim}
\end{equation}
For $n=\omega/2$, $\alpha=0$ and $\beta=-1$ and $\theta=2\rho$, the form of the asymptotic modes can be written as
\begin{equation}
\Phi_{\omega_n}=\sqrt{\frac{1}{2\pi}\frac{2\cot{\rho}}{\pi\omega_n}}e^{-i\omega_n t}\cos\left(\omega_n\rho-\frac{\pi}{4}\right).
\label{Aphin2}
\end{equation}
 The modes are required to be orthonormal under the norm (\ref{scalarproduct}) which can be written as
\begin{equation}
(\Phi_1,\Phi_2)=i\int_{\Sigma}d^dx\sqrt{g}g^{tt}(\Phi_1^*(\partial_t\Phi_2)-(\partial_t\Phi_1^*)\Phi_2).
\label{Ascalarproduct}
\end{equation}
 Taking the positive frequency solution and the negative frequency solution of equation (\ref{Aphin2}) and calculating the norm in the high frequency limit $(\omega_n=2n)\ n,m\gg 1$, we get
\begin{equation}
\int_0^{\frac{\pi}{2}}\int_0^{2\pi}\frac{1}{2\pi}\frac{4n}{\sqrt{\pi n\pi m}}\tan(\rho)\sqrt{\cot(\rho)}\sqrt{\cot(\rho)}\cos(2n\rho-\frac{\pi}{4})\cos(2m\rho-\frac{\pi}{4})d\rho d\theta=\delta_{mn}.
\label{Aorthonormal}
\end{equation}
The scalar field in the collapse geometry, where the ingoing null rays become outgoing null rays after being reflected at the center, can then be expanded in terms of the mode functions
\begin{equation}
\Phi=\sum_{\omega_{n}} \left(a_{\omega_n}\Phi_n +a^{\dagger}_{\omega_n}\bar{\Phi}_n\right)
\label{Afadsmodeexp}
\end{equation}
where the modes are given by the standard form in the collapse setting,
\begin{equation} 
\Phi_n =\sqrt{\frac{1}{2\pi}\frac{2}{\pi\omega_n}}(\cot(\rho))^{1/2}(e^{-i\omega_nF(V)}-e^{-i\omega_nU}),
\label{Afasymodes3}
\end{equation}
where $U,V$ are the outgoing and ingoing rays defined by $U=t-\rho$ and $V=t+\rho$. Computing $<0|T_{t\rho}|0>$ using the modes of the asymptotic form (\ref{Afasymodes3}) where $\rho=\pi/2$ (putting $\sin \pi/2=1$) gives,
\begin{equation}
<0|T_{t\rho}|0>=\frac{1}{2\pi^2}\cos(\rho)\sum_{\omega_n}\left[\omega_n\left(F'(V)F'(V+\epsilon)e^{i\omega_n[F(V+\epsilon)-F(V])}-e^{i\omega \epsilon}\right)\right]
\label{Afordparker1}
\end{equation}

\begin{equation}
+\frac{1}{\pi^2}(1-F'(V))\sum_{\omega_n}\sin(\omega_n(F(V)-U))
\label{Afordparker2}
\end{equation}
where $\epsilon$ is introduced for regularization. The expression is split into two parts (\ref{Afordparker1}) and (\ref{Afordparker2}). The 
expression (\ref{Afordparker2}) vanishes because in the collapsing solution the map of rays
satisfies $U=F(V)$.
The expression (\ref{Afordparker1}) is a sum over even integers. In the high frequency limit the sum can be replaced by an integral. Evaluating the integral and integrating over the disk of radius $\tan\rho$ (at $\rho=\pi/2$)
gives the flux of particles. The quantum flux is evaluated to give
\begin{equation}
\int_o^{2\pi}<0|T_{\rho t}|0>\tan(\rho)d\theta=\frac{1}{12\pi}\left[\frac{F'''}{F^{'3}}-\frac{3}{2}\left(\frac{F''}{F^{'2}}\right)^2\right].
\label{Afordparker5}
\end{equation}
For simplicity, we have worked out only the case $m=0$ ($e^{im\phi}=1$), and as before, carried
out a point splitting regularization. Written in terms of the inverse
function $V=G(U)$ the expression (\ref{Afordparker5}) for the radiated power
becomes
\begin{equation}
P=\int<0|T_{\rho t}|0>Rd\phi=\frac{1}{2\pi }\left[\frac{1}{4}\left(\frac{G''}{G'}\right)^2-\frac{1}{6}\frac{G'''}{G'}\right]
\label{fpformulaPa}
\end{equation}
which should be compared with the expression for the radiated power
(\ref{fpformulaP}) in the flat case. The two expressions are essentially
identical, except for a factor of two.

\section {Null rays for collapse on an AdS background}
The interior metric in the AdS case is given by (\ref{nmetricref})
 and the BTZ exterior by (\ref{nexteriorref}). In this section we evaluate the formula required to derive the map between the ingoing and outgoing null rays. In the exterior, the null geodesic equation is given by
$V=T+R^*$ and $U=T-R^*$, where $V$ is an ingoing null ray and $U$ is an outgoing null ray. $R^*$ is the tortoise coordinate defined as, 
\begin{equation}
dR^*=\frac{dR}{\Lambda R^2-M}.
\label{nrstar}
\end{equation}
For the case $M<0$ we have,
\begin{equation}
R^*=\frac{\tan^{-1}(\sqrt{\Lambda /m}R)}{\sqrt{m\Lambda}}
\label{nnMrstar}
\end{equation}
where $m=-M$. For the case $M=0$ we have
\begin{equation}
R^*=-\frac{1}{\Lambda R}.
\label{nzMrstar}
\end{equation}
For the case $M>0$ we have
\begin{equation}
R^*=\frac{\log{\left|\left({\frac{\sqrt{\Lambda} R-\sqrt{M}}{\sqrt{\Lambda} R+\sqrt{M}}}\right)\right|}}{2\sqrt{M\Lambda}}.
\label{npMrstar}
\end{equation}
 The exterior time $T$ can be expressed in terms of the interior time $t$ by matching the first fundamental form. Comparing the coefficients of $d\phi^2$ for the metrics (\ref{nmetricref}) and (\ref{nexteriorref}) yields
\begin{equation}
R=r_0\cos(\sqrt{\Lambda}t)+B\sin{\Lambda}t),
\label{nqR1}
\end{equation}
where $r_0$ is the outer boundary of the dust cloud. Comparing the remaining coefficients of the first fundamental form gives,
\begin{equation}
-dt^2=-(\Lambda R^2-M)dT^2+\frac{dR^2}{(\Lambda R^2-M)}.
\label{nqtTexterior}
\end{equation} 

The equation (\ref{nqR1}) can be differentiated with respect to $t$ to obtain $\dot{R}$.
This gives 
\begin{equation}
\frac{dT}{dt}=\frac{\sqrt{\Lambda r_0^2+\Lambda B^2-M}}{\Lambda R^2-M}=\frac{q}{\Lambda (r_0\cos(\sqrt{\Lambda}t)+B\sin(\sqrt{\Lambda}t))^2-M}
\label{nqdTdtB}
\end{equation}
where $q=\sqrt{\Lambda r_0^2+\Lambda B^2-M}$. 
The above equation is solved for the three cases.
For $M>0$ we get
\begin{equation}
T=\frac{1}{2\sqrt{\Lambda M}}\ln{\frac{|q+\sqrt{M}\tan(t')|}{|q-\sqrt{M}\tan(t')|}}.
\label{nTtMp}
\end{equation}
For $M=0$ we have
\begin{equation}
T=\frac{\tan(t')}{\sqrt{\Lambda q}} +c.
\label{nTtMz}
\end{equation}
For $M<0$ we define $m=-M$, and get
\begin{equation}
T=\frac{1}{\sqrt{\Lambda m}}\arctan\left(\frac{\sqrt{m}\tan(t')}{q}\right)
\label{nTtMn}
\end{equation}
where $t'=\sqrt{\Lambda} t+s$ and $s=\arccos ({{r_0}/{\sqrt{r_0^2+b^2}}})$.

For the interior, the radial null geodesic equation is obtained by imposing the null condition on the metric (\ref{nmetricref}). We take the initial velocity $B(r)$ and the initial density $\rho_0(r)$ to be constants $b,2k$ respectively.
We get
\begin{equation}
dt\sec(\sqrt{\Lambda} t)=\pm \frac{dr}{\sqrt{\Lambda r^2 +\Lambda b^2- kr^2+c}}
\label{nnullgeodb1}
\end{equation}
Solving the above equation we can show the interior ingoing null ray $v$ to be,
\begin{equation}
v=\frac{1}{{\sqrt{\Lambda}}}\ln{|\sec(\sqrt{\Lambda} t)+\tan(\sqrt{\Lambda} t)|}+\frac{1}{\sqrt{d}}\sinh^{-1}{\sqrt{\frac{d}{a}}r}
\label{nnullintv}
\end{equation}
and the interior outgoing null ray $u$ is
\begin{equation}
u=\frac{1}{{\sqrt{\Lambda}}}\ln{|\sec(\sqrt{\Lambda} t)+\tan(\sqrt{\Lambda} t)|}-\frac{1}{\sqrt{d}}\sinh^{-1}{\sqrt{\frac{d}{a}}r}
\label{nnullintu}
\end{equation}
where $a=\Lambda b^2+c$ and $d=(\Lambda-k)r^2$.

\section{Particle creation for the naked case $M<0$}
For the case $M<0$, we have the relation between the interior and the exterior time coordinates, 
given by (\ref{nTtMn})
\begin{equation}
T=\frac{1}{\sqrt{\Lambda m}}\arctan\left(\frac{\sqrt{m}\tan(t')}{q}\right)
\label{nTtMn2}
\end{equation}
where $t'=\sqrt{\Lambda} t+s$ and $s=\arccos\left(\frac{r_0}{\sqrt{r_0^2+b^2}}\right)$. And the outer shell is given by $R_0(t')=\sqrt{r^{2}_0+b^2}cos( t')$.
The interior null rays are given by (\ref{nnullintv}) and (\ref{nnullintu}). The exterior null rays are given by $V=T+R^*$ and $U=T-R^*$ where $R^*$ is the tortoise coordinate given by (\ref{nnMrstar}). 
Singularity forms when $t=0$ for the central shell $r=0$. 
Let the ingoing null ray in the interior coordinates which gets mapped onto the Cauchy horizon be $v_0$. Putting $t=0,r=0$ in equation (\ref{nnullintv}) we get the equation for $v_0$ to be
\begin{equation}
v_0=0=\frac{\ln{|\sec(\sqrt{\Lambda} t)+\tan(\sqrt{\Lambda} t)|}}{\sqrt{\Lambda}}+ \frac{1}{\sqrt{d}}\sinh^{-1}{\sqrt{\frac{d}{a}}r}.
\label{nnullintv3}
\end{equation}
The time $t_0$ at which this null ray entered the dust cloud is given by putting the value of $r$ to be the outer shell radius $r_0$ and solving for $t_0$. From this $T$ can be computed using equation (\ref{nTtMn2}). The exterior $R$ can be computed using the relation $R_0(t')=\sqrt{r^{2}_0+b^2}\cos( t')$. So $V_0$ (the null ray in the exterior which finally becomes the Cauchy horizon) can be evaluated. Now the functions 
$T(t'),\ R_0(t')$ are differentiable functions in the domain under consideration, so the incoming null rays can be Taylor expanded about $V_0$ (which gets mapped onto $v_0$). The mapping in the neighborhood of $V_0$ is $V=pv+i$ ($p,i$ are constants) since $\dot{T}(t')$ and $\dot{R_0}(t')$ are approximately constant. A dot
denotes derivative with respect to $t$.
At the center the ingoing null rays get mapped onto outgoing null rays with the linear relation $v=u$. Let the Cauchy horizon be $u_0$. When the interior outgoing null ray reaches the outer shell, it becomes the outgoing null ray in the exterior. At the boundary, the relation between the exterior and interior outgoing null rays becomes $U=wu+y$ where $w,y$ are constants.
So the map $U=F(V)$ between $V$ and $U$ is linear even near the Cauchy horizon.
For a linear map, the equation (\ref{Afordparker5}) implies that the particle flux is zero since the formula involves second and third derivatives of the map $U=F(V)$. This is
unlike in the case of 3+1 spherical dust collapse, where the emitted 
quantum flux diverges on the Cauchy horizon.

\section{Particle creation for the black hole case $M>0$}

For the case $M>0$ we have
\begin{equation}
R^*=\frac{\log{|({\frac{\sqrt{\Lambda} R-\sqrt{M}}{\sqrt{\Lambda} R+\sqrt{M}}})|}}{2\sqrt{M\Lambda}}.
\label{npMrstar2}
\end{equation}
$T$ in terms of the interior time $t$ is given by
\begin{equation}
T=\frac{1}{2\sqrt{\Lambda M}}\ln{\frac{|q+\sqrt{M}\tan(t')|}{|q-\sqrt{M}\tan(t')|}}.
\label{nTtMp2}
\end{equation}
Null geodesics are given by $U=T-R^*$ and $V=T+R^*$.
The interior null geodesics are given by equations (\ref{nnullintv}) and (\ref{nnullintu}).
We shall consider the case where a null ray $V_0$  enters the dust cloud to become $v_0$. It gets mapped onto a outgoing null ray $u_0$ which forms the event horizon. We analyze the map of the null rays which enter the dust cloud just before the the null ray $V_0$ enters. 
The event horizon forms when the outer shell's physical radius becomes $R_h=\sqrt{M/\Lambda}$. Expressed in terms of the interior time, the physical radius of the outer shell is $R_h= r_0\cos(\sqrt{\Lambda }t)+b\sin(\sqrt{\Lambda }t)=\sqrt{r_0^2+b^2}\cos(t')$ where $t'=\sqrt{\Lambda} t+s$.
So $\cos(t')=\sqrt{\frac{M}{\Lambda(r_0^2+b^2)}}$.
The mapping around $V_0$ when the null ray enters the outer edge will be linear since the functions $R^*$ and $T$ are well behaved at the sufficiently earlier epoch when there is no event horizon formed. So $V=pv+f$. $p,f$ are constants. The mapping at the center of the cloud is linear $u=v$. The mapping from $u$ to $U$ is non-trivial because $T$ in equation (\ref{nTtMp2}) tends to infinity since $q=\sqrt{M}\tan(t')$ at the horizon. Similarly $R^*$ tends to infinity since $\Lambda R^2-M$ goes to zero. Let $t_0$ be the interior time when the event horizon gets formed; we Taylor expand the expressions inside the logarithms. For $R$ tending to $R_h$ we get

\begin{equation}
T=\frac{-1}{2\sqrt{\Lambda M}}\ln{\left|\frac{\Lambda(r_0^2+b^2)(t'_0-t')}{\sqrt{M}}\right|}
\label{nTtMp4}
\end{equation}
and $R^*$ is given in the limit $R$ tending to $R_h$ by
\begin{equation}
R^*=\frac{1}{2\sqrt{\Lambda M}}\ln{|q(t'_0-t')|}.
\label{nTtMpR4}
\end{equation}
So $U$, which is given by $U=T-R^*$, becomes
 \begin{equation}
U-U_0=\frac{\ln{|y(t_0-t)|}}{\sqrt{\Lambda M}}
\label{hawk}
\end{equation}
where $y=\sqrt{\frac{q\Lambda (r_0^2+b^2)}{\sqrt{M}}}.$
Now $t_0-t=a(V_0-V)+h$ since the mapping is linear till the null ray comes out of the outer edge. Hence the map for $V<V_0$ is,
\begin{equation}
  U-U_0=F(V)=\frac{\ln{|y(V_0-V)|}}{\sqrt{\Lambda M}}.        
\label{hawk1}
\end{equation}

Following the method in Section 7 we derive the Hawking flux 
in our model of 2+1 circular dust collapse, using the map of incoming and outgoing null rays in the asymptotic region, found in Eqn. (\ref{hawk1}) above.
Substituting the map (\ref{hawk1}) in equation (\ref{Afasymodes3}) we get

\begin{equation}
\Phi_n =\sqrt{\frac{1}{2\pi}\frac{2}{\pi\omega_n}\cot(\rho)}(e^{-i\omega_n\frac{1}{\sqrt{\Lambda M}}\ln(y[V-V_0])}-e^{-i\omega_nU}).
\label{asymodes5}
\end{equation}
For the given modes, we can compute the total power radiated using the formula (\ref{Afordparker5});
we get $P={\Lambda M}/{24\pi}$.The rate of absorption of radiation by a blackhole immersed in a thermal bath of temperature $T$ is given by $\pi T^2/6$ in (2+1) dimensions \cite{bsdewitt}.
Equating the two expressions for power the temperature $T$ of the 
blackhole is found to be  
\begin{equation}
T_h=\frac{\sqrt{\Lambda M}}{2\pi}=\frac{\kappa}{2\pi}
\label{temperature}
\end{equation}
where $\kappa$ is the surface gravity of the blackhole \cite{birrel}.

The temperature can be obtained also by computing the Bogoliubov transformation between the modes of the 'in' region and the 'out' region. The modes in the 'in' region and the 'out' region can respectively be taken to be of the form \cite{s5},\cite{fordnotes},
\begin{equation}
\Phi_n =\sqrt{\frac{1}{2\pi}\frac{2}{\pi\omega_n}\cot(\rho)}(e^{-i\omega_nV}),
\label{bogoinmode}
\end{equation}

\begin{equation}
{\Psi}_m =\sqrt{\frac{1}{2\pi}\frac{2}{\pi\omega_m}\cot(\rho)}(e^{-i\omega_m(F(V)}).
\label{bogooutmode}
\end{equation}
The Bogoliubov coefficients are given by the expressions
\begin{equation}
\alpha_{\omega_n\omega_m}=(\Phi_n,\Psi_m),
\label{alpa}
\end{equation}
\begin{equation}
\beta_{\omega_n\omega_m}=-(\Phi_n,\Psi^{*}_m),
\label{beta}
\end{equation}
where the bracket is the norm defined in (\ref{Ascalarproduct}).
The mean particle number with the frequency $\omega_m$ is given by $\sum_n |\beta_{\omega_n\omega_m}|^2$.
The expression for (\ref{beta}), upon integrating over the angular variable gives,
\begin{equation}
\beta_{\omega_n\omega_m}=\frac{4}{\pi}\sqrt{\frac{\omega_n}{\omega_m}}\int^{V_0}_{-\infty}dVe^{-i\omega_nV}e^{-i\omega_mF(V)}.
\label{beta2}
\end{equation} 
The equation (\ref{beta2}) is evaluated for the map (\ref{hawk1}). It is a standard integral \cite{fordnotes} and can be shown to yield a Planckian spectrum with the mean number of particles given by,
\begin{equation}
N_{\omega_n}=\sum_{\omega_m}|\beta_{\omega_n\omega _m}|^2=\frac{1}{{e^{\frac{2\pi}{\sqrt{\Lambda M}}\omega_n}}-1}
\label{plankian}
\end{equation}
Hence the temperature can be obtained to be $T=\frac{\sqrt{\Lambda M}}{2\pi}$,
which agrees with the temperature deduced above from the radiated power.

\section{Concluding remarks}
Unlike in the case of 3+1 dust collapse, no particle creation takes place
when a naked singularity forms in 2+1 collapse. This suggests that the 
quantum gravitational investigation of singularity avoidance in the 2+1 case
may be more tractable compared to the 3+1 case. Further, these results 
were found using the Ford-Parker formula that we derived afresh for the 2+1 
case. Since the Ford-Parker method used here yields the expected
Hawking radiation when the collapse ends in a BTZ black hole, the approach
can be assumed to be reliable and robust. In a forthcoming work, we will 
build upon the results presented here to set up a canonical quantum 
gravitational treatment of 2+1 circularly symmetric dust collapse.

Nonetheless, one should also inquire about the physical reason(s) which
make the result in the 2+1 dimensional spacetime different from that in
the 3+1 case. The reason for this difference is the presence of a negative 
cosmological constant in the 2+1 case. This constant softens
the radiated power due to the way in which an incoming wave
is blueshifted and then, as it turns outgoing, is redshifted
along the Cauchy horizon. The difference also arises because the singularities in the 2+1 case are conical in nature. In a future work we also plan to investigate the role of dimensionality - if one studies quantum particle creation in
a higher dimensional AdS spacetime admitting a naked singularity, is the
radiated power zero or not?

The occurrence of a naked singularity in gravitational collapse in an AdS 
spacetime is possibly generic, and allowed also in four dimensions, as 
discussed in \cite{hhm1}, \cite{hhm2} and \cite{mann2}. In \cite{hhm1} 
the gravitational collapse of a scalar field with a potential $V(\phi)$
in an 4-d AdS spacetime was considered. It was then numerically shown
that for a large class of potentials an open set of initial data evolves to
naked singularities in asymptotically AdS solutions.
In \cite{hhm2} asymptotically AdS solutions in N=8 supergravity
having a negative total energy were examined. Some of these negative energy 
solutions were shown to contain classical evolution of regular initial data
leading to naked singularities. These numerical results are supported by
the analytical work of \cite{mann2} and \cite{sashi}.

These results possibly have very
important consequences for the AdS/CFT correspondence, which maps a
`bulk gravity theory' in an AdS spacetime to a quantum conformal field theory
on the AdS boundary. Will this correspondence survive if the gravity theory
generically admits a naked singularity? This significant issue remains open,
and deserves to be investigated carefully; it has been argued in \cite{hhm2}
that the AdS/CFT correspondence will imply the absence of singularities
in a quantum theory of gravity. It is also possible that the occurrence of
a naked singularity could compel us to redefine the classical gravity theory 
and/or the quantum conformal field theory in the AdS/CFT correspondence.

\bigskip

\bigskip

\noindent{\bf Acknowledgments}: It is a pleasure to thank Claus Kiefer, Cenalo Vaz and Rakesh Tibrewala for useful discussions.


\begin{thebibliography}{99}
\bibitem{si1} T. P. Singh, in 
{\it Classical and quantum aspects of gravitation and cosmology,}
Eds. G. Date and B. R. Iyer (1996), [gr-qc/9606016].
\bibitem{si2} T. P. Singh, in {\it General Relativity and Gravitation} Ed. M. Sasaki, J. Yokoyama, T. Nakamura and K. Tomita; Osaka University (2001)
[gr-qc/0012087].
\bibitem{ford} L. H. Ford and L. Parker, Phys. Rev D17, 1485 (1978).
\bibitem{hiscock} W. A. Hiscock, L. G. Williams and D. M. Eardley,
 Phys. Rev. D26, 751 (1982). 
\bibitem{vaz1} C. Vaz and L. Witten, Class. Quant. Grav. 12 (1995) 2607.
\bibitem{vaz2} C. Vaz and L. Witten, Nucl. Phys. B487 (1997) 409.
\bibitem{s1} S. Barve, T. P. Singh. C. Vaz and L. Witten,
Nuclear Physics B 532, 361 (1998).
\bibitem{s2} S. Barve, T. P. Singh. C. Vaz and L. Witten,  Phys.
Rev. D58, 104018 (1998).
\bibitem{s3} S. Barve, T. P. Singh. and C. Vaz,
Phys. Rev. D62 (2000) 084021. 
\bibitem{s4} T. P. Singh and C. Vaz,  Phys. Lett. B481 (2000) 74.
\bibitem{s5} T. P. Singh and C. Vaz,  Phys. Rev. D61 (2000) 124005.
\bibitem{sj1} P. S. Joshi and T. P. Singh, Phys. Rev. D51,
6778 (1995).
\bibitem{sj2} T. P. Singh and P. S. Joshi, Classical and Quantum Gravity 13, 559 (1996).
\bibitem{sj3} S. Jhingan, P. S. Joshi and T. P. Singh,  Classical
and Quantum Gravity 13, 3057 (1996).
\bibitem{six} T. Harada, H. Iguchi, K. Nakao, T. P. Singh, T. Tanaka and 
C. Vaz, Phys. Rev. D64 (2001) 041501.
\bibitem{wd} C. Vaz, L. Witten and T. P. Singh, Phys.Rev.D63 (2001) 104020.
\bibitem{vawi1} C. Vaz and L. Witten, Phys. Rev. D60 (1999) 024009.
\bibitem{vawi2}C. Vaz and L. Witten, Phys. Rev. D63 (2001) 024008.
\bibitem{k1} C. Vaz, C. Kiefer, T. P. Singh and L. Witten,
 Phys. Rev. D67 (2003) 024014.
\bibitem{k2} C. Vaz, L. Witten and T. P. Singh, Phys. Rev. D69 (2004) 104029. 
\bibitem{k3} C. Kiefer, J. Mueller-Hill and C. Vaz, Phys. Rev. D73, 044025 (2006).
\bibitem{k4} C. Kiefer, J. Mueller-Hill, T. P. Singh and C. Vaz, gr-qc/0703008.
\bibitem{sashi} S. Gutti, Class. Quantum Grav. 22, 3223 (2005).  
\bibitem{mann1} S.F.Ross, R.B. Mann, Phys.Rev.D47:3319-3322 (1993), hep-th/9208036.
\bibitem{mann2} R.B. Mann, John J.Oh, Phys.Rev.D74:124016 (2006), gr-qc/0609094  
\bibitem{peleg} Y. Peleg, A. R. Steif, Phys. Rev. D51, 3992 (1995).
\bibitem{barrow} J.D.Barrow, D.J.Shaw, C.G.Tsagas, Class.Quant.Grav.23:5291-5322, 2006, gr-qc/0606025.
\bibitem{hortacsu1} T.Birkandan, M. Hortacsu, Gen.Rel.Grav.35:457-466,2003, gr-qc/0104096.
\bibitem{hortacsu2} M. Hortacsu, H.T.Ozcelik, B.Yapiskan, Gen.Rel.Grav.35:1209-1221,2003, gr-qc/0302005
\bibitem{banados} M. Banados, C.Teitelboim and J. Zanelli, Phys. Rev. Lett. 69, 1849 (1992).
\bibitem{isham} S.J. Avis, C. J. Isham and D. Storey, 
Phys. Rev. D18, 3565 (1978).
\bibitem{balasubramaniyan} V.Balasubramaniyan, P. Kraus, 
A. Lawrence, Phys. Rev. D59, 046003 (1999).
\bibitem{vakkuri} S. Hemming, E. Keski-Vakkuri, Phys.Rev.D, Vol 64, 044006 (2001).
\bibitem{korea} Seungjoon Hyun, Geon Hyoung Lee, Jae Hyung Yee, Phys.Lett.B 322 (1994) 182.
\bibitem{CGHS} C. G. Callan, Jr., S. B. Giddings, J. A. Harvey and 
A. Strominger, Phys. Rev. D45 (1992) 1005.
\bibitem{GKS} N. Goheer, M. Kleban and L. Susskind, JHEP 0307:066 (2003).
\bibitem{Witten} E. Witten, arxiv:0706.3359 [hep-th].
\bibitem{hawking} S. W. Hawking and G. F. R. Ellis,
{\it The Large Scale Structure of Spacetime} (Cambridge, 1971).
\bibitem{table} {\it Table of Integrals, Series and Products}, 
I.S. Gradshteyn and I.M. Ryzhik. 
\bibitem{birrel} N.D. Birrel and P. C. W Davies, 
{\it Quantum Fields in Curved Space},
(Cambridge (1982)).
\bibitem{bsdewitt} B.S. DeWitt, Phys.Rept.19:295-357, 1975.
\bibitem{fordnotes} L.H. Ford, {\it Quantum Field theory in Curved Spacetime}, gr-qc/9707062.
\bibitem{hhm1} T. Hertog, G. T. Horowitz and K. Maeda, 
Phys. Rev. Lett. 92 (2004) 131101.
\bibitem{hhm2} T. Hertog, G. T. Horowitz and K. Maeda, 
Phys. Rev. D69 (2004) 105001.

\end{thebibliography}
\end{document}